\documentclass[a4paper,12pt]{article}
\usepackage{amssymb,amsmath}
\usepackage{bm}

\def\beq{\begin{equation}}
\def\eeq{\end{equation}}
\def\bea{\begin{eqnarray}}
\def\eea{\end{eqnarray}}
\def\nn{\nonumber}
\def\bZ{{\mathbb Z}}
\def\cF#1{{\cal F}_{#1}}
\def\VS{Vect($S^1$)}
\def\g{{\mathfrak g}_{\ell}}
\def\sg{{\mathfrak {sg}}_{\ell}}

\def\hf{\frac{1}{2}}
\begin{document}
%
%
%
%
\thispagestyle{empty}
\vspace*{3cm}
\begin{center}
\textbf{\Large Aspects of infinite dimensional $\ell$-super Galilean conformal algebra} \\

\vspace{10mm}
{\large
N. Aizawa${}^1$ and J. Segar${}^2$
}

\bigskip
1. Department of Physical Science, 
Graduate School of Science, 
Osaka Prefecture University, 
Nakamozu Campus, Sakai, Osaka 599-8531, Japan

\medskip
2. Department of Physics, 
Ramakrishna Mission Vivekananda College, 
Mylapore, Chennai 600 004, India

\end{center}

\vskip 3cm
\begin{abstract}
In this work we construct a infinite dimensional $\ell$-super Galilean conformal algebra, which is a generalization of the 
$\ell=1$ algebra found in the literature. 
We give a classification of central extensions, 
the vector field representation, the coadjoint representation and the operator
product expansion of the infinite dimensional $\ell$-super Galilean conformal algebra, keeping possible applications in physics 
and mathematics in mind.  
\end{abstract}

\clearpage
\setcounter{page}{1}
%
%
%
\section{Introduction}

 The Virasoro algebra, which is an infinite dimensional Lie algebra, is found to play 
a very vital role in many areas of physics and mathematics. 
Various aspects of this algebra such as coadjoint representations, supersymmetric extensions, 
orbits, vertex operators are well studied in the literatures. 
 
 A simple generalization of the Virasoro algebra is constructed by taking semidirect sum of 
the Virasoro algebra and an abelian ideal:
\bea
  [L_n, L_m] &=& (n-m) L_{n+m} + c_1 n(n^2-1) \delta_{n+m,0},
  \nn \\[3pt]
  \nn [L_n, P_m] &=& (n-m) P_{n+m} + c_2 n(n^2-1) \delta_{n+m,0},
  \\[3pt]
  [P_n, P_m] &=& 0, \qquad n, m \in {\mathbb Z} \label{leq1}
\eea
This infinite dimensional Lie algebra appears in various context of physical and mathematical problems in recent years.

In general relativity, the algebra is called $\mathfrak{bms}_3 $ and describes an infinite dimensional symmetry of 
three-dimensional asymptotically flat spacetimes at null infinity \cite{AsBiSch,Bagchi,BarCom,BarGomGon,BarOb1,BarOb2,DuGibHo} 
(and references therein and see \cite{BMS4,Sachs} for original four-dimensional version). 
The same algebra appears in non-relativistic analogue of AdS/CFT correspondence and called 
infinite dimensional Galilean 
conformal algebra (GCA) \cite{MarTachi,BaGoMaMi,HoRo,HeSchStoUn,HosNas,HenHosRou} 
(see also \cite{BagGop,AlDaVah,HoRo2,Henkel,RogUnter,AndBolo} for some varieties of the algebra). 
The same algebra also appears in cosmological topologically massive gravity \cite{HoKuNi,Bagchi} and bosonic string theory \cite{Bagchi3,BaChaPar}. 
It is shown that the Navier-Stokes equation  can be recast in a form covariant under GCA \cite{Muk} 
(see also \cite{ChernHen,ZhangHov} for GCA covariant equations). 
It is also shown that GCA appears as a symmetry of Newton-Cartan spacetime \cite{DuvHov}. 

 The GCAs we mentioned are all infinite dimensional and have the Virasoro algebra as a subalgebra.  
These can be regarded as an infinite dimensional extension of the GCAs of \textit{finite} dimension 
which were found in \cite{HavPre,HenkelPRL,NdORM1}. The structure of the finite dimensional GCAs is a semidirect sum of $ sl(2,\mathbb{R}) \oplus so(d) $ and 
an abelian ideal.  The Virasoro sector of infinite dimensional GCAs is replaced by its subalgebra $ sl(2,{\mathbb R}). $ 
The finite dimensional GCAs also attract much attention recently in various contexts. 
A nice review of physical and mathematical aspects of GCAs is found in \cite{RogUnter} (see the forward of it) 
and we quote some very recent works \cite{Mas1,Chern,AKT1,AKT2,KriLechSo,ChernGala}. 

 We now return to the algebra (\ref{leq1}). The Virasoro algebra is the central extension of the Lie algebra of vector fields on $ S^1. $ 
From this viewpoint one can consider extensions of the Virasoro algebra by modules of tensor densities on $ S^1. $ 
The algebra (\ref{leq1}) was obtained as one of such extensions \cite{OvRo,OvRo2} and 
its relations to matrix Strum-Liouville operators were investigated \cite{Mar}. 
The algebra (\ref{leq1}) is called $W(2,2)$ in the context of vertex operator algebras. 
The $W(2,2)$ algebra and its highest weight modules arise naturally in the studies of vertex operator algebra $ L(\hf,0) \otimes L(\hf,0) $ 
\cite{ZhangDong, JiaPei, Rado}.

 However, in the context of infinite dimensional GCA, the algebra (\ref{leq1}) is a special case of a wider class of 
infinite dimensional Lie algebras. This class of Lie algebras was introduced in \cite{MarTachi,Muk} and  
each member of the class  is labelled by two parameters $(d,\ell)$ where $ d $ takes any positive integer and 
$ \ell $ takes a non-negative integer or a non-negative half-integer value. 
Its structure is a semidirect sum of Virasoro $\oplus \widehat{so}(d) $ and infinite dimensional abelian ideal. 
The algebra (\ref{leq1}) is $ d = \ell = 1 $ member  of this class of Lie algebras. 
From mathematical point of view it is quite natural not to restrict ourselves to a specific case such as $ d=\ell=1$, 
but to study the class of Lie algebras itself.  
However, higher values of $d$ require additional generators so that structure of the algebra is altered a lot. 
Therefore, as a first step, we keep $ d = 1 $ and restrict ourselves to  this subclass of Lie algebras for a generic $\ell $ in the present work. 

For a given value of $ \ell $ the infinite dimensional Lie algebra 
that we investigate is defined by  the commutation relations:
\bea
  [L_n, L_m] &=& (n-m) L_{n+m} + c_1 n(n^2-1) \delta_{m+n,0},
  \nn \\[3pt]
  \nn [L_n, P_r] &=& (\ell n-r) P_{n+r} + c_2 n(n^2-1) \delta_{n+r,0} \delta_{\ell,1},
  \\[3pt]
  [P_r, P_s] &=& 0, \qquad n, m \in {\mathbb Z}, \ r, s \in {\mathbb Z} + \ell \label{AlgDef}
\eea
The suffix $r$ of $P_r$ takes an integer or a half-integer value depending on the value of $ \ell. $ 
It is known that $c_1$ and $ c_2 $ exhaust all possible central extensions \cite{OvRo2,Hosse}. 
We denote this algebra by $\g$ throughout this paper.

 On the other hand, supersymmetry is a very fundamental notion in physics and in mathematics. 
It is therefore natural to consider extensions of our algebras (\ref{AlgDef}) to superalgebras. 
Supersymmetric extensions found in literatures are also restricted to $ \ell = 1 $ algebra. 
Such extensions are considered again in the context of infinite dimensional GCA \cite{BagMan,Mandal,HoPlyVal}, BMS group \cite{AwaGibShaw,BaDoMaTro,Banks},   
generalized Sturm-Liouville operators \cite{Mar2,MaOvsRog} and string theory \cite{ManRay,Mandal2}. 
We introduce a supersymmetric extension of the algebra (\ref{AlgDef}) for any values of $\ell,$ 
and classify its central extensions. 
This is one of the purposes of the present work. 
Supersymmetric extensions of the finite dimensional GCAs are also studied in a lot of literatures. 
For example, see \cite{BeHu,DuvHor,DuGiHo,SaYo1,SaYo2,Saka,HenUnter,HoPlyVal,Naru1,Naru2,MasSuper}.

  Having a class of infinite dimensional Lie algebras and its supersymmetric extension, 
we aim to provide some basic tools that will open ways to further physical and mathematical applications of this algebraic structure. 
Among others,  as a first step, we focus on vector field representations, coadjoint representations 
and operator product expansions (OPE) in the present work. 

  The present paper is organized as follows. 
In \S \ref{SEC:2} the bosonic Lie algebra (\ref{AlgDef}) is extended to $ {\cal N} =1 $ superalgebra and we list 
up all possible central extensions of the  superalgebra. 
In \S \ref{SEC:3} vector field representation of the superalgebra without central extensions is given. 
Based on this vector field representation we reconstruct the whole superalgebra in terms of tensor densities of 
the Lie algebra of a smooth vector field on $ S^1. $ 
This allows us to define a regular dual of the  superalgebra which is used to construct the coadjoint representation. 
In \S \ref{SEC:4} following the standard procedures we construct the operator product expansion for the superalgebra. 
We close the paper with some concluding remarks in \S \ref{SEC:5}.

\setcounter{equation}{0}
\section{Supersymmetric extension and central extensions}
\label{SEC:2}
\subsection{$\ell$-Super GCA for $d=1$}
It is well known that the supersymmetric algebras play a very fundamental role in many areas of physics and mathematics. 
Realising the importance of the supersymmetry, we want to extend the algebra defined in (\ref{AlgDef}) to a superalgebra. 
We look for the minimal extension to the supersymmetric case. We could construct a supersymmetric algebra, with two additional fermionic 
generators, namely $G_m$ and $H_r$, where $ m \in {\mathbb Z}, \ r \in {\mathbb Z} + \ell. $ 
The following relations define a Lie superalgebra:
\bea
  & & [L_m, L_n] =(m-n) L_{m+n} + c_1 m(m^2-1) \delta_{m+n,0},
  \nn \\
  & & [L_m, P_r] =(\ell m-r) P_{m+r} + c_2 m(m^2-1) \delta_{m+r,0} \,\delta_{\ell, 1},
  \nn \\
  & & [P_r, P_s] = 0, \nn \\
  & & \{ G_m, G_n \} = 2 L_{m+n} + c_1(4m^2-1) \delta_{m+n,0},
  \nn \\
  & & \{ G_m, H_r \} = 2P_{m+r} + c_2(4m^2-1) \delta_{m+r,0} \,\delta_{\ell, 1},
  \nn \\
  & & [L_m, G_n] = \Big( \frac{m}{2}-n \Big) G_{m+n}, 
  \nn \\
  & & [L_m, H_r] = \Big( \frac{2\ell-1}{2}m -r \Big) H_{m+r},
  \nn \\
  & & [P_r, G_m] = \Big( \frac{r}{2}-\ell m\Big) H_{r+m},
  \nn \\
  & & [P_r, H_s] = \{H_r, H_s \} = 0, \label{l-SUSY-def}
\eea
where $ m, n \in {\mathbb Z} $ and $ r, s \in {\mathbb Z} + \ell. $ 
Here this algebra is written down with all possible central extensions. 
This classification of central extensions will be proved in the next subsection. 
The supersymmetric extension (\ref{l-SUSY-def}) recovers the algebra introduced in \cite{Mandal} for $\ell=1.$ 
The subset spanned by $  L_m $ and $ G_m  $ is the Ramond algebra which is a ${\cal N} = 1 $ extension of the Virasoro algebra. 
One may construct the superalgebra that has the Neveu-Schwarz as a subalgebra by adding two 
fermionic generators $ G_r, H_m \ (r \in {\mathbb Z} + \ell, m \in {\mathbb Z}) $ to (\ref{AlgDef}). 
The results of the following sections may not be altered a lot for this second $\ell$-super GCA. 
Therefore we consider only the algebra (\ref{l-SUSY-def}) in this work. 
The superalgebra defined by (\ref{l-SUSY-def}) is denoted by $\sg.$ 

 Different types of supersymmetric extension of $ \ell = 1 $ GCA are introduced and some aspects of them are 
discussed in \cite{BagMan,BaDoMaTro,Banks,Mar2,MaOvsRog,ManRay,Mandal2}.

%
\subsection{Proof of the central extensions}

  The central extension of the bosonic sector of (\ref{l-SUSY-def}) has been classified in \cite{Hosse}. 
We thus consider the following additional central extensions:
\bea
  & & \{ G_m, G_n \} = 2 L_{m+n} + Z_{mn},
  \nn \\
  & & \{ G_m, H_r \} = 2P_{m+r} + \alpha_{mr},
  \nn \\
  & & [L_m, G_n] = \Big( \frac{m}{2}-n \Big) G_{m+n} + \beta_{mn}, 
  \nn \\
  & & [L_m, H_r] = \Big( \frac{2\ell-1}{2}m -r \Big) H_{m+r} + \gamma_{mr},
  \nn \\
  & & [P_r, G_m] = \Big( \frac{r}{2}-\ell m\Big) H_{r+m} + \xi_{rm},
  \nn \\
  & & [P_r, H_s] = \eta_{rs}, \qquad \{H_r, H_s \} = W_{rs}, \label{CentralExtensions}
\eea
where $ Z_{mn} = Z_{nm}, W_{rs} = W_{sr}, \alpha_{mr}, \beta_{mn}, \gamma_{mr}, \xi_{rm} $ and $ \eta_{rs} $ 
are abelian generators that commute with all other generators. 
We shall use the super-Jacobi identities in order to single out all possible central extensions.  

  The super-Jacobi identity for $\{ P_r, H_s, H_t \}$ yields the equation:
\beq
   \eta_{rt} H_s + \eta_{rs} H_t = 0.
\eeq
It follows that $ \eta_{rs} =0 $ for all $r,s.$ 
For $\{ P_r, G_m, H_s \} $ we obtain the equation:
\beq
    \xi_{rm} H_s = 0,
\eeq
which shows that $ \xi_{rm} = 0. $ 
For $ \{ P_r, L_m, G_n \} $ we have the equation:
\beq
   \Big( \frac{r}{2} -\ell n \Big) \gamma_{m,r+n} = 0.
\eeq
Setting $ n = 0 $ one sees that $ \gamma_{mr} = 0 $ if $ r \neq 0. $ 
While setting $ r+n = 0 \ (r^2+n^2 \neq 0) $ one see that $\gamma_{m,0} = 0. $ 
Therefore we showed that $ \gamma_{mr} = 0  $ for all $m$ and $r.$

  We have two independent relations from the super-Jacobi identity for $\{ L_m, G_n, H_r \}:$
\bea
  & & \beta_{mn} H_r = 0,
  \label{LGH1} \\[3pt]
  & & \Big( \frac{2\ell-1}{2}m -r \Big) \alpha_{n,m+r} + \Big(\frac{m}{2}-n \Big) \alpha_{m+n,r} 
      = 2c_2 m(m^2-1) \delta_{m+n+r,0} \delta_{\ell,1}.
  \label{LGH2}
\eea
The equation (\ref{LGH1}) shows that $ \beta_{mn} = 0. $ 
From (\ref{LGH2}) with $ m=0 $  we have
\beq
   (r+n) \alpha_{nr} = 0. 
\eeq
Hence $ \alpha_{nr} = 0 $ if $ \ell $ is a half-integer since $ r+n \neq 0 $ for this case. 
If $ \ell $ is an integer, one may write $ \alpha_{nr} = \delta_{n+r,0} a_n $ and (\ref{LGH2}) 
yields the equation:
\beq
   \Big( \frac{2\ell+1}{2}m +n \Big) a_n + \Big(\frac{m}{2}-n \Big) a_{m+n} 
      = 2c_2 m(m^2-1) \delta_{\ell,1} 
   \label{LGH3}
\eeq
for all values of $m, n$. The equation (\ref{LGH3}) with $ m=1$ gives a recurrence relation for $ a_n:$
\beq
   \Big( \ell+n+\hf \Big) a_n + \Big(\hf -n \Big) a_{n+1} = 0
\eeq
which is solved to give the formula:
\beq
  a_n = - \frac{ (2\ell+2n-1)!! }{ (2n-3)!! (2\ell-1)!! } a_0.
\eeq
The equation (\ref{LGH3}) with $ m=-n$ gives an another expression of $a_n:$
\beq
  a_n = \frac{3}{1-2\ell} a_0 - \frac{4 c_2 (n^2-1) }{1-2\ell}\delta_{\ell,1}.
\eeq
From these two formulae for $ a_n$ one may see that
\beq
    a_0 = \left\{
      \begin{array}{cl}
        -c_2 & \ell = 1 \\[5pt]
        0 & \ell \neq 1
      \end{array}
    \right.
\eeq
It follows that 
\beq
    a_n = \left\{
      \begin{array}{cl}
        c_2 (4n^2-1) & \ell = 1 \\[5pt]
        0 & \ell \neq 1
      \end{array}
    \right.
\eeq
Hence we obtain
\beq
   \alpha_{mn} = c_2 (4m^2 - 1) \delta_{m+n,0} \delta_{\ell,1}. 
\eeq

 Next we show that $ W_{rs} = 0. $ The super-Jacobi identity for $\{ L_m, H_r, H_s \} $ gives the equation:
\beq
  \Big( \frac{2\ell-1}{2}m - s \Big) W_{r,m+s} + \Big( \frac{2\ell-1}{2}m - r \Big) W_{s,m+r} = 0. \label{LHH1}
\eeq
By setting  $ m=0, $ we have that 
$
  (r+s) W_{rs} = 0. 
$ 
This implies $ W_{rs} = \delta_{r+s,0}\, \omega_r. $ 
Therefore (\ref{LHH1}) yields
\beq
   \Big( \ell m + \frac{m}{2} +r \Big) \omega_r + \Big( \ell m - \frac{m}{2} -r \Big) \omega_{m+r} = 0, 
   \label{LHH2}
\eeq
for all $ m, r. $ From (\ref{LHH2}) with $ m= -2r $ we obtain $ \ell r \omega_r = 0. $ 
Therefore $ \omega_r = 0 $ if $ r \neq 0. $ Note that this is always true for half-integer $\ell$. 
We need to show that $ \omega_0 = 0 $ for integer $ \ell. $ 
From (\ref{LHH2}) with $ r = 0 $ and non-vanishing $m$ we have 
\[
  \Big( \ell + \hf \Big) \omega_0 + \Big( \ell - \hf \Big) \omega_m = 0. 
\]  
It follows that $ \omega_0 = 0 $ as $ \omega_m = 0 $ for $m \neq 0. $ 
Thus we showed $ W_{rs} = 0. $  

 Finally, we show that $ Z_{mn} $ gives a non-trivial central extension. 
It follows from the super-Jacobi identity for $ \{ L_k, G_m, G_n \} $ which gives the equation:
\beq
  \Big( \frac{k}{2} -n \Big) Z_{m,n+k} + \Big( \frac{k}{2} -m \Big) Z_{n,m+k} = 2 c_1 k (k^2-1) \delta_{m+n+k,0}. 
  \label{LGG1} 
\eeq
One may see that $ Z_{mn} = \delta_{m+n,0} z_m $ since we have $ (m+n) Z_{mn} = 0 $ from the equation 
(\ref{LGG1}) with $ k =0.$ With this expression of $ Z_{mn} $ (\ref{LGG1}) is written as follows:
\beq
  \Big( \frac{3k}{2} + m \Big) z_m + \Big( \frac{k}{2} - m \Big) z_{m+k} = 2 c_1 k(k^2-1),
  \label{LGG2}
\eeq
for all $ m, k. $ Form the equation (\ref{LGG2}) with $ k = 2m $ we obtain that 
$ m z_m = c_1m (4m^2 -1) $  which implies that
\beq
  z_m = c_1 (4m^2-1) \quad \text{for} \ m \neq 0  \label{LGG3}
\eeq 
To obtain $z_0$ we use the equation (\ref{LGG2}) with $ k = -m \ (\neq 0):$
\beq
  z_m + 3z_0 = 4c_1 (m^2-1). 
\eeq
This together with (\ref{LGG3}) gives that $ z_0 = -c_1. $ Hence we obtain 
$ z_m = c_1 (4m^2-1) $ for all $m$ and $ Z_{mn} = \delta_{m+n,0}\, c_1 (4m^2-1). $ 

 This completes the proof of the classification of all possible central extensions.

%
%
%
%
\setcounter{equation}{0}
\section{Some representations}
\label{SEC:3}

\subsection{Vector field representation}

The vector field representations are explicit realizations of the infinitesimal actions of the generators of the group. 
Having an explicit representation helps us in doing practical calculations, for instance, constructing invariant equations of the group. 
We write down the vector field representation of the $\ell$-super GCA without central extensions which  
is  given in terms of one Grassmann variable: 
\bea
L_m &=& -t^{m+1} \partial_t - (m+1) t^m \left( \ell x \partial_x + \frac{1}{2} \xi \partial_{\xi} \right), \nn \\
P_r&=&- \frac{1}{2}t^{r+\ell}\partial_x, \nn \\
G_m &=&-t^{m +\frac{1}{2}} (\xi \partial_t - \partial_{\xi}) - (2m+1) \ell  t^{m-\frac{1}{2}} \xi x\partial_{x}, \nn \\
H_{r}&=&-t^{r+ \ell -\frac{1}{2}} \xi \partial_x, \label{l-SUSY-rep1}
\eea
where $t,x$ are $c$-number variables and $ \xi $ is a Grassmann variable.

  We remark that a realization for the $ \ell = 1 $ algebra with two Grassmann variables is known in the literature \cite{Mandal}:
\bea
L_n &=& -t^{n+1} \partial_t - (n+1)t^n x\partial_x -\frac{1}{2}(n+1)\left[t^n(\alpha\partial_{\alpha}+
\beta\partial_{\beta}) + nt^{n-1}x\alpha\partial_{\beta}\right], \nn \\
P_n&=&-t^{n+1}\partial_x -\frac{1}{2}(n+1)t^n\alpha\partial_{\beta}, \nn \\
G_m &=&-t^{m+ \frac{1}{2}}(\alpha\partial_t + \beta\partial_x -\partial_{\alpha})
-(m+\frac{1}{2})t^{m-\frac{1}{2}}x(\alpha\partial_x-\partial_{\beta}), \nn \\
H_{r}&=&-t^{r+\frac{1}{2}}(\partial_{\beta} -\alpha\partial_x), \label{l-1-rep}
\eea
where $ \alpha, \beta $ are Grassmann variables. 

  The representations (\ref{l-SUSY-rep1}) and (\ref{l-1-rep}) also give representations of the bosonic algebra $\g.$ 
By restricting to bosonic generators we have a representation in terms of $c$-number and Grassmann variables. 
If we further remove the Grassmann variables, we have a representation in terms only of $c$-number variables.  

  Now we consider the current algebra corresponding to (\ref{l-SUSY-rep1}). 
Let $ f(t), g_{\ell}(t), \gamma(t) $ and $ \chi_{\ell}(t)$ be $C^{\infty}$ functions of $t$.  
Then the representation (\ref{l-SUSY-rep1}) is a Laurent expansion of the followings:
\bea
  & & L_f = f(t) \partial_t + \ell f'(t) x \partial_x + \frac{1}{2} f'(t) \xi \partial_{\xi},
    \nn \\
  & & P_{g_{\ell}} = \hf g_{\ell}(t) \partial_x, \nn \\
  & & G_{\gamma} = \gamma(t) (\xi \partial_t - \partial_{\xi})  + 2\ell \gamma'(t) \xi x \partial_x,
    \nn \\
  & & H_{\chi_{\ell}} = \chi_{\ell}(t) \xi \partial_x. \label{l-SUSY-rep2}
\eea
Indeed, by expanding the functions as below one may recover (\ref{l-SUSY-rep1}):
\bea
  & & f(t) = -\sum_{n \in \bZ} c_n t^{n+1}, 
      \qquad 
      g_{\ell}(t) = -\sum_{n \in \bZ} \alpha_n t^n = - \sum_r \alpha_r t^{\ell+r},
  \nn \\
  & & \gamma(t) = -\sum_{n \in \bZ} \beta_n t^{n+\hf}, 
     \qquad 
     \chi_{\ell}(t) = - \sum_r \sigma_r t^{r+\ell-\hf}.
     \label{LaurentExp}
\eea
More concretely, we have
\beq
  L_f = \sum_n c_n L_n, \qquad P_{g_{\ell}} = \sum_r \alpha_r P_r,
  \qquad 
  G_{\gamma} = \sum_n \beta_n G_n, \qquad H_{\chi_{\ell}} = \sum_r \sigma_r H_r.
  \label{LaurentExp2}
\eeq
The non-vanishing commutators for (\ref{l-SUSY-rep2}) are given by
\bea
  & & [L_f, L_g] = L_{fg'-f'g}, \qquad\qquad  [L_f, P_{g_{\ell}}] = P_{fg'_{\ell}- \ell f'g_{\ell}},
  \nn \\[3pt]
  & & [L_f, G_{\gamma}] = G_{f\gamma'- \hf f'\gamma}, \qquad \quad
      [L_f, H_{\chi_{\ell}}] = H_{ f \chi'_{\ell}- \frac{2\ell-1}{2} f' \chi_{\ell} },
  \nn \\[3pt]
  & & [P_{g_{\ell}}, G_{\gamma}] = H_{ \ell g_{\ell} \gamma' - \hf g'_{\ell} \gamma },  
      \qquad \ 
      \{ G_{\gamma}, H_{\chi_{\ell}} \} = P_{-2\gamma \chi_{\ell}},
  \nn \\[3pt]
  & & \{ G_{\gamma}, G_{\delta} \} = L_{-2\gamma \delta}. \label{CurrentSuAlg}
\eea
These commutation relations help us in  
identifying the generators with tensor densities of Vect$(S^1)$  
which is considered in the next section.

%
\subsection{Tensor density module construction of $\sg$ and regular dual}

 From now on we compactify the variable $ t= e^{i\theta}. $ 
Therefore the functions  $ f(t), g_{\ell}(t), \gamma(t) $ and $ \chi_{\ell}(t)$ are defined on $S^1.$ 
 Let Vect$(S^1)$ be the Lie algebra of a smooth vector field on $ S^1:$ 
$f(t) d/dt.$  
Let $ \cF{\lambda} $ be the space of all tensor densities on $S^1$ of degree $\lambda : \phi = \phi(t) (dt)^{-\lambda}. $ 
The Lie algebra Vect$(S^1)$ acts on $ \cF{\lambda} $ by the Lie derivative:
\beq
  L_{f(t)\frac{d}{dt}} \phi(t) dt^{-\lambda} = (f\phi' - \lambda f' \phi) dt^{-\lambda}.
\eeq
This leads to the identification of the elements of the superalgebra to the tensor density module:
\beq
  L_f \simeq \cF{1}, \qquad P_{g_{\ell}} \simeq \cF{\ell}, \qquad 
  G_{\gamma} \simeq \cF{\hf}, \qquad H_{\chi_{\ell}} \simeq \cF{\ell-\hf}.
  \label{TenDenIdentification}
\eeq
Indeed, by an appropriate definition of (anti)commutation relations for $ \cF{\lambda} $ one may define a Lie 
superalgebra for the vector space of the tensor density modules. 
Furthermore the Lie superalgebra is isomorphic to the one defined in (\ref{CurrentSuAlg}). 
First, we  note the isomorphism 
\beq
 \text{\VS}  \ltimes \cF{\ell} \simeq \langle \;  L_f, \ P_{g_{\ell}} \; \rangle,
  \label{S1F-direct}
\eeq
where $ \ltimes $ denotes the semidirect sum. 
The isomorphism is established by the following commutation relation for \VS $ \ltimes \cF{\ell}:$
\beq
 [ (f,g_{\ell}), (h, p_{\ell})] = (fh' - f'h, fp'_{\ell}-\ell f' p_{\ell} - h g'_{\ell} + \ell h' g_{\ell}).
 \label{CMdirectSum}
\eeq
 Next, we consider the \VS $ \ltimes \cF{\ell} $ module $ M_{\ell} = \cF{\hf} \oplus \cF{\ell-\hf}. $ 
We define the action $T_{(f, g_{\ell})}$ of \VS $ \ltimes \cF{\ell} $ on $ M_{\ell} $ by 
\beq
  T_{(f, g_{\ell})} \begin{pmatrix} \gamma dt^{-\hf} \\ \chi_{\ell} dt^{-(\ell-\hf)} \end{pmatrix} 
  =
  \begin{pmatrix}
     (f\gamma' - \hf f' \gamma) dt^{-\hf} \\
    ( f \chi'_{\ell} - (\ell -\hf) f' \chi_{\ell} + \ell g_{\ell} \gamma' - \hf g'_{\ell} \gamma )
    dt^{-(\ell-\hf)}
  \end{pmatrix}.
  \label{TactM}
\eeq
In commutator form
\beq
  [(f, g_{\ell}), (\gamma, \chi_{\ell})]
  = \left(f\gamma' - \hf f' \gamma,\; f \chi'_{\ell} - (\ell -\hf) f' \chi_{\ell} + \ell g_{\ell} \gamma' - \hf g'_{\ell} \gamma \right).
  \label{TactM-CM}
\eeq
We define the anticommutator $ M_{\ell} \otimes M_{\ell} \to $ \VS $ \ltimes \cF{\ell}$ by the relation:
\beq
 \{\, (\gamma, \chi_{\ell}), (\delta,\sigma_{\ell}) \,\} 
 = (-2\gamma \delta, -2(\gamma \sigma_{\ell}+ \delta \chi_{\ell}) ).
 \label{AntiCommM}
\eeq
Then it is immediate to verify the followings: 
(i) equations (\ref{CMdirectSum}), (\ref{TactM-CM}) and (\ref{AntiCommM}) define a Lie superalgebra structure on  \VS $\oplus \cF{\ell} \oplus M_{\ell}. $ 
(ii) the constructed Lie superalgebra is isomorphic to the one defined by (\ref{CurrentSuAlg}).

  Now we incorporate the central extension. They are given by
\bea
  & & c_1(L_f, L_g) =  \int_{S^1} f'(t) g''(t) dt, 
      \qquad
      c_1(G_{\gamma}, G_{\delta}) =  \int_{S^1} \gamma'(t) \delta'(t) dt,
  \nn \\[3pt]
  & & c_2(L_f, P_{g_{\ell}}) = \delta_{\ell 1}\,  \int_{S^1} f'(t) g''_{\ell}(t) dt, 
      \quad
      c_2(G_{\gamma}, H_{\chi_{\ell}}) = \delta_{\ell 1}\,\int_{S^1} \gamma'(t) \chi'_{\ell}(t) dt.
    \nn\\
  & & \label{centers}
\eea
The Virasoro central extension is the well-known Gelfand-Fuchs cocycle. 
In this way we have achieved the tensor density construction of the superalgebra $\sg:$
\beq
  \sg = \cF{1} \oplus \cF{\ell} \oplus \cF{\hf} \oplus \cF{\ell-\hf} \oplus {\mathbb R} \oplus \delta_{\ell 1} 
  {\mathbb R},
  \label{modulestrLieSA}
\eeq
where $ \oplus $ denotes the direct sum of vector spaces. 
 
In order to discuss coadjoint representations, we would like to introduce the algebraic dual of $ \sg. $  
As usual in the infinite dimensional setting, we consider the regular dual of $ \sg $ (see \cite{Kirillov}). 
Here, specifically, the dual module $ \cF{\lambda}^* $ may be identified with $ \cF{-\lambda-1} $ through the pairing:
\beq
 \langle \varphi (dt)^{\lambda+1}, \phi (dt)^{-\lambda} \rangle 
 = 
 \int_{S^1} \varphi(t) \phi(t) dt. 
\eeq
We thus identify the dual algebra $ \sg^*$ as follows:
\beq
  \sg^* = \cF{-2} \oplus \cF{-\ell-1} \oplus \cF{-\frac{3}{2}} \oplus \cF{-\ell-\hf} \oplus {\mathbb R} \oplus \delta_{\ell 1} 
  {\mathbb R}.
  \label{RegDualLieSA}
\eeq

\subsection{Coadjoint representation of $\sg$}
A Lie algebra acts on itself by the adjoint action provided by the Lie bracket. 
Given a Lie algebra and its dual, we can find the coadjoint action of the algebra on its dual.
The coadjoint action $ ad^* $ of $ \sg $ on  $\sg^*$ is defined by
\beq
  \langle ad^* X (a), Z \rangle = \langle a, [Z, X]_{\pm} \rangle, \quad 
  X, Z \in \sg, \ a \in \sg^*
  \label{CoAd-def} 
\eeq
where $[ \ , \ ]_{\pm} $ denotes the commutator or anticommutator and 
the duality pairing between $ \sg $ and $ \sg^* $ is defined by
\bea
  & & \langle \, \vec{x}, \vec{X} \, \rangle =
  \int_{S^1} ( \alpha f + \beta_{\ell} g_{\ell} + a\gamma + b_{\ell} \chi_{\ell}) dt + 
  \sum_{i=1}^2 \kappa_i c_i,
  \nn \\
  & & \vec{x} = ( \alpha, \beta_{\ell}, a, b_{\ell}, \kappa_1, \kappa_2) \in \g^*
  \nn \\
  & & \vec{X} = (f, g_{\ell}, \gamma, \chi_{\ell}, c_1, c_2) \in \g
  \label{DualPairing}
\eea
Using the algebra $\sg$ and the pairing relation, it may not be difficult to derive the following 
formulae, here the action on the central elements is omitted since it is trivial: 
\beq
  ad^* L_f(\vec{x}) = 
  \begin{pmatrix}
     f \alpha' + 2f' \alpha - \frac{i\kappa_1}{2\pi} f''' \\[3pt]
     f \beta'_{\ell} + (\ell+1) f' \beta_{\ell} - \delta_{\ell 1} \frac{i\kappa_2}{2\pi} f''' \\[3pt]
     f a' + \frac{3}{2} f' a \\[3pt]
     f b'_{\ell} + (\ell+\hf) f' b_{\ell}
  \end{pmatrix},
  \label{cadL}
\eeq
\beq
  ad^* P_{g_{\ell}}(\vec{x}) = 
  \begin{pmatrix}
     \ell g_{\ell} \beta'_{\ell} + (\ell+1) g'_{\ell} \beta_{\ell} - \delta_{\ell 1} \frac{i\kappa_2}{2\pi} g'''_{\ell} \\[3pt]
     0 \\[3pt]
     \ell g_{\ell} b'_{\ell} + (\ell+\hf) g'_{\ell} b_{\ell} \\[3pt]
     0
  \end{pmatrix}
  \label{cadP}
\eeq
\beq
  ad^* G_{\gamma}(\vec{x}) =
  \begin{pmatrix}
    \hf \gamma a' + \frac{3}{2} \gamma' a  \\[3pt]
    \hf \gamma b'_{\ell} + (\ell + \hf) \gamma' b_{\ell}\\[3pt]
    -2\gamma \alpha - \frac{2i\kappa_1}{\pi} \gamma'' \\[3pt]
    -2 \gamma \beta_{\ell} - \delta_{\ell 1} \frac{2i\kappa_2}{\pi} \gamma''
  \end{pmatrix} 
  \label{cadG}
\eeq
\beq
  ad^* H_{\chi_{\ell}}(\vec{x}) = 
  \begin{pmatrix}
     (\ell-\hf) \chi_{\ell} b'_{\ell} + (\ell+\hf) \chi'_{\ell} b_{\ell}   \\[3pt]
     0  \\[3pt]
     -2 \beta_{\ell}  \chi_{\ell} - \delta_{\ell 1} \frac{2i\kappa_2}{\pi} \chi''_{\ell} \\[3pt]
     0
  \end{pmatrix}
  \label{cadH}
\eeq
If we restrict ourselves to the bosonic coordinates only and take the Grassmann variables to zero we get the
coadjoint representation for the algebra $\g$. 
With this coadjoint representation one may consider the coadjoint orbit of the (super)groups generated by $\g$ or $ \sg. $ 
The coadjoint orbits of the BMS${}_3 $ group (which is generated by ${\mathfrak g}_1$) is already discussed in \cite{BarOb1} using the induced representations. 
We also mention seminal works on coadjoint orbits of the Virasoro and super Virasoro groups \cite{Segal,Witten,Bakas}.

%
%
%
%
\setcounter{equation}{0}
\section{OPE for the algebras}
\label{SEC:4}

In this section we construct the operator product expansion for fields which can be expanded in terms of the generators 
of (\ref{l-SUSY-def}) . 
We follow the standard approach used in the theory of vertex operator algebras (see for example \cite{Kac,Schotten}).
For illustrative purpose we start with the bosonic part of the algebra.
We write down the fields corresponding to $L_n$ and $P_r$ as power series in the variable $z:$ 
\bea
L(z) &=& \sum_{n \in \mathbb{Z}} L_n z^{-n-2} \\
P(z) &=& \sum_{r \in \mathbb{Z} + \ell} P_r z^{-r-\ell -1} \label{LPexpansion}
\eea
The OPE of $L(z)L(w)$ is well-known:
\beq
   L(z) L(w) \sim \frac{6c_1}{(z-w)^4} + \frac{2 L(w)}{(z-w)^2} + \frac{\partial_w L(w)}{z-w}.
\eeq
To find the OPE of $L(z)$ and $P(w)$, we calculate
\bea
 [L(z), P(w) ] &=& \sum_{m,r} [L_m , P_r ] z^{-m-2} w^{-r-\ell -1} \nn \\
  &=& \sum_{m,r} (\ell m-r) P_{m+r} z^{-m-2} w^{-r-\ell-1} \nn \\
  & & \quad 
  + \, \delta_{\ell,1} c_2 \sum_m m(m^2-1) z^{-m-2} w^{m-2} \label{LzPw}
\eea
Now consider the first term, setting $s=m+r$ then $n=m+1$:
\bea
 & & \sum_{m,r} (\ell m - r) P_{m+r} z^{-m-2} w^{-r- \ell-1} 
 \nn \\
 & & \qquad 
   = (\ell+1) \sum_{n,s} (P_s w^{-s-\ell-1})\, \partial_w z^{-n-1} w^n 
   + \sum_{n,s} (P_s \partial_w w^{-s-\ell-1}) z^{-n-1} w^n
 \nn \\
 & & \qquad 
   = (\ell+1) P(w) \partial_w \delta(z-w) + (\partial_w P(w)) \delta(z-w).
\eea
The definition of the formal delta function $ \delta(z-w) = \sum_n z^{-n-1} w^n $ was used in the last equality. 
The second term in (\ref{LzPw}) is easily calculated by setting $n=m+1$:
\beq
  \sum_m m(m^2-1) z^{-m-2} w^{m-2} =
  \sum_n z^{-n-1} \partial_w^3\, w^n = \partial_w^3\, \delta(z-w).
\eeq
We thus have 
\beq
 [L(z), P(w) ] = (\ell+1) P(w) \partial_w \delta(z-w) + (\partial_w P(w)) \delta(z-w) 
   + \delta_{\ell,1} c_2\,  \partial_w^3\, \delta(z-w).
\eeq
Due to the equivalence between 
\beq
[A(z), B(w) ] = \sum_{j=0}^{N-1} C^j (w) \frac{1}{j!}\, \partial_w^j \,\delta(z-w) 
\eeq
and 
\beq
A(z)B(w) \sim \sum_{j=0}^{N-1} \frac{C^j(w)}{(z-w)^{j+1}} 
\eeq
we obtain
\beq
L(z)P(w) \sim  \frac{6 c_2\, \delta_{\ell,1}}{(z-w)^4} + \frac{(\ell+1) P(w)}{(z-w)^2}+ \frac{ \partial_w P(w) }{(z-w)}
\label{OPE2}
\eeq
It is now obvious that OPE for the product of two $ P(z) $ is given by
\beq
P(z)P(w) \sim 0.
\eeq

Next we work out the operator product expansion for $\sg.$ 
The fields corresponding to the generators of this algebra  are expanded as (\ref{LPexpansion}) 
and as:
\beq
G(z) = \sum_{n \in {\mathbb{Z}}} G_n z^{-n -\frac{3}{2}} \qquad H(z) = \sum_{r \in \mathbb{Z} +\ell} H_r z^{-r -\ell -\frac{1}{2}}.
\eeq
OPE of two fermionic fields is computed in a way similar to the above example of $ L(z) P(z)$ provided that 
the commutator is replaced with anticommutator. 
It may not be difficult to verify the following OPE:
\bea
L(z)G(w)  &\sim&   \frac{3}{2}\frac{G(w)}{(z-w)^2} + \frac{\partial_{w} G(w)}{(z-w)} \nonumber \\[3pt]
G(z)H(w) &\sim&  \frac{8 c_2\, \delta_{\ell,1}}{(z-w)^3}+ \frac{2P(w)}{(z-w)} \nonumber \\[3pt]
L(z)H(w) &\sim& \Big(\ell + \frac{1}{2} \Big) \frac{H(w)}{(z-w)^2}+\frac{\partial_{w} H(w)}{(z-w)} \nonumber \\[3pt]
P(z)G(w) &\sim& \Big(\ell+\frac{1}{2} \Big) \frac{H(w)}{(z-w)^2} + \ell\, \frac{\partial_{w} H(w)}{(z-w)}  \nonumber \\[3pt]
G(z)G(w) & \sim &  \frac{8c_1}{(z-w)^3} +  \frac{2L(w)}{(z-w)}  \nonumber \\[3pt]
P(z)H(w) & \sim & 0 \nonumber \\[3pt]
H(z)H(w) & \sim &0
\eea
In principle it is possible to calculate various correlation functions for theories with primary field
which realize $\sg$ as a symmetry algebra. 
Readers may refer the recent works \cite{HenSym,HenSto} (and references therein) for a detailed study of two-point functions.

\section{Concluding remarks }
\label{SEC:5}

In this paper we have considered an infinite dimensional $\ell$-super Galilean conformal algebra of $d=1.$ 
We gave a classification of central extensions and 
have worked out some of the basic ingredients of representation theories such as vector representation, tensor density module construction,
coadjoint representation and operator product expansion for this superalgebra. The physical and mathematical
applications of the superalgebra and the structures that are worked out  here will be taken up
for investigations in the future.

  We restricted ourselves to $d=1$ algebras which is the simplest subclass of the larger class of $ \ell$-GCA. 
Representation theories and supersymmetric extensions of $\ell$-GCA for higher values of $d$ have not been studied extensively. 
As found in the literatures \cite{BagMan,BaDoMaTro,Banks,Mar2,MaOvsRog,ManRay,Mandal2}, supersymmetric extension of 
$\g$ is not unique and some of them are discussed in connection with tensionless string theory. 
This implies that starting from bosonic $\ell$-GCA for higher values of $d$ one may obtain 
various infinite dimensional superalgebras and some of them are of physical interest. 
Therefore further studies of $\ell$-(super) GCA will  provide us fruitful results in both 
physics and mathematics. 

\subsection*{Acknowledgement} J. S would like to thank Professor Naruhiko Aizawa, for invitation and hospitality at OPU and the 
organizers of Workshop on Vertex Operators and Mock Modular Forms, NUI, Galway, Ireland, for hospitality. 
N. A. is supported by the  grants-in-aid from JSPS (Contract No. 26400209).

%
%
%

\end{document}